\documentclass[ms,11pt]{egs}       
  \usepackage{times}                   

\usepackage{graphicx}

\printfigures              

\begin{document}
\title{Geomagnetic control of the spectrum of traveling ionospheric
       disturbances based on data from a global GPS network}

%
%
%
%
\author[1,*]{E. L. Afraimovich}
\author[1]{E. A. Kosogorov}
\author[1]{O. S. Lesyuta}
\author[1]{I. I. Ushakov}
\affil[*]{p. o. box 4026, Irkutsk, 664033, Russia\\
          fax: +7 3952 462557; e-mail: afra@iszf.irk.ru\\}
\affil[1]{Institute of Solar-Terrestrial Physics SD RAS}
\author[2]{A. F. Yakovets}
\affil[2]{Institute of Ionosphere, Almaty, Kazakhstan}
%
%
\date{Manuscript version from 3 December 2000}

\journal{Annales Geophysicae}       
%
%
\firstauthor{Afraimovich} \proofs{E. L. Afraimovich\\Institute of
Solar-Terrestrial Physics SD RAS\\ Irkutsk, Russia} \offsets{E. L.
Afraimovich\\Institute of Solar-Terrestrial Physics SD RAS\\
Irkutsk, Russia}

\msnumber{12345}

\received{28 August 2000}
\revised{15 September 2000} 
\accepted{3 December 2000}

\runninghead{Smith and Weston: Winning the West}
\firstpage{123}
\pubyear{2001}
\pubvol{25}
\pubnum{2}

\maketitle

%
%
%

\begin{abstract}
In this paper an attempt is made to verify the hypothesis on the
role of geomagnetic disturbances as a factor determining the
intensity of traveling ionospheric disturbances (TIDs). To
improve the statistical validity of the data, we have used the
based on the new GLOBDET technology \citep{Afr2000a} method
involving a global spatial averaging of disturbance spectra of
the total electron content (TEC). To characterize the TID
intensity quantitatively, we suggest that a new global index of
the degree of disturbance should be used, which is equal to the
mean value of the rms variations in TEC within the selected range
of spectral periods (of 20--60 min in the present case). The
analysis has been made for a set of 100 to 300 GPS stations, and
for 10 days with a different level of geomagnetic activity ($Dst$
from 0 to -350 nT; the $Kp$ index from 3 to 9).

It was found that power spectra of daytime TEC variations in the
range of 20--60 min periods under quiet conditions have a
power-law form, with the slope index \mbox{$k$ = -2{.}5}. With an
increase of the level of magnetic disturbance, there is an
increase in total intensity of TIDs, with a concurrent kink of
the spectrum caused by an increase in oscillation intensity in
the range of 20--60 min. The TEC variation amplitude is found to
be smaller at night than during the daytime, and the spectrum
decreases in slope, which is indicative of a disproportionate
increase in the amplitude of the small-scale part of the spectrum.

It was found that an increase in the level of geomagnetic
activity is accompanied by an increase in total intensity of TEC;
however, it correlates not with the absolute level of $Dst$, but
with the value of the time derivative of $Dst$ (a maximum
correlation coefficient reaches -0{.}94). The delay of the TID
response of the order of 2 hours is consistent with the view that
TIDs are generated in auroral regions, and propagate equatorward
with the velocity of about 300-400 m/s.
\end{abstract}

\section*{Keywords}
Ionospheric disturbances $\cdot$ Auroral ionosphere $\cdot$
Equatorial ionosphere

\section{Introduction}\label{sec:intro}

One of the most important ideas of the origin of ionospheric
irregularities and their dynamics is the wave concept, according
to which the experimentally observed irregular structure is the
result of the superposition of wave processes of different
origins. For that reason, particular significance in theoretical
and experimental studies has been attached to spectral
characteristics which make it possible to identify ionospheric
irregularities of different scales.

The irregularities discussed in this paper are classed as
traveling ionospheric disturbances (TIDs), with a typical spatial
size ranging from 100 to 1000 km, and a time period in the range
of 20--120 min, and the literature on this subject is quite
extensive. A classification of TIDs in their sizes (in
particular, their separation into large-scale LS and medium-scale
MS disturbances) is rather arbitrary, and many authors attribute
different physical mechanisms to this classification.

LS TIDs with typical time periods of 1--2 hours and wavelengths of
about 1000 km have been studied in many publications, including
some thorough reviews \citep{Hun82,Hoc96}. It is commonly
accepted that LS TIDs are manifestations of acoustic-gravity
waves (AGW) whose generating regions are located in the auroral
zones of the northern and southern hemispheres. These ideas have
been confirmed in recent experiments using GPS data \citep{Ho98,
Afr2000b}.

There is currently a plethora of views regarding the
effectiveness of geomagnetic field disturbances as the source of
medium-scale (MS) TIDs. According to Hunsucker (1982), the
auroral source plays the dominant role for electron density
irregularities with typical periods from 10 to 60 min. An
enhancement of the variation intensity during geomagnetic
disturbances was pointed out by F\"orster et al. (1994) and
Fatkullin et al. (1996).

At the same time Waldock and Jones (1987) showed that the auroral
sources perhaps play a minor role in the generation of MS TIDs
recorded at mid-latitudes. Ogawa et al. (1987) hold that MS TIDs
are constantly recorded based on observations from the NNSS
satellites, and their occurrence frequency does not increase
under disturbed conditions. A plausible mechanism for the
production of MS TIDs is attributed by some authors to
meteorological processes \citep{Bert75, Wal87, Oli97}.

On this basis, one is led to conclude that there has been as yet
no sufficiently convincing evidence in support of the hypothesis
of the determining contribution of geomagnetic disturbances to
the production of medium--scale TIDs. This is largely caused by
the inadequate number and low spatial resolution of the radio
sounding facilities currently in use (ionosondes, incoherent
scatter radars, etc.).

The main characteristics of wave processes are the temporal and
spatial spectra. Since the spectra have normally a power-law
character, the slope of the spectrum $k$ and the standard
deviation of intensity variations in the frequency range $M$
analyzed (the amplitude scale of the power spectrum) are the most
informative parameters, estimates of which were made in almost all
publications of an experimental or theoretical nature
\citep{Dro79, Lit83, Kali88, Fri90, Yak99}.

Determining the above-mentioned characteristics of disturbances
experimentally is of crucial importance for validating the
interpretation of experimental data in terms of different
physical mechanisms of the inhomogeneous structure. Furthermore,
a knowledge of irregularity spectra is required for developing an
empirical model of distortions of transionospheric signals used
in special--purpose radio engineering systems of communication,
location, and navigation in the meter, decimeter and centimeter
ranges.

Published data show a large scatter in estimates of the slope $k$
and of the amplitude scale $M$ of temporal and spatial spectra
(see Section 4). One of the reasons for this scatter might be that
different measuring techniques are used, which differ greatly in
spatial and temporal resolution. However, the main reason is
determined by the differing geophysical conditions of separate
measurements, and by the large difference in latitude, longitude
and local time when carrying out experiments.

To obtain more reliable information requires carrying out
simultaneous measurements over a large area covering regions with
a different local time. However, none of the above-mentioned
methods meets such requirements.

The advent and evolution of a Global Positioning System, GPS, and
also the creation on its basis of widely branched networks of GPS
stations (at least 757 sites by November of 2000, the data from
which are placed on the Internet), opened up a new era in remote
ionospheric sensing \citep{Klo97}. High-precision measurements of
the group and phase delay along the line of sight (LOS) between
the receiver on the ground and transmitters on the GPS system
satellites covering the reception zone are made using
two-frequency multichannel receivers of the GPS system at almost
any point of the globe and at any time simultaneously at two
coherently coupled frequencies $f_1=1575{.}42$ MHz and
$f_2=1227{.}60$ MHz.

One of these authors \citep{Afr2000a} has developed a new
technology, GLOBDET, for global detection of ionospheric
disturbances of natural and technogenic origins using data from
the international network of two-frequency multichannel receivers
of the navigation GPS system which improves substantially the
sensitivity and spatial resolution of experiment.

The objective of this study is to develop, on the basis of the
GLOBDET technology, a new method for estimating global
characteristics of the TID spectrum which excels in a higher
statistical reliability that is achieved by a global spatial
averaging of the spectra. This method is used to verify the
hypothesis of the determining role of geomagnetic disturbances as
the source of TIDs.

The geometry and a general description of experiments are given
in Section 2. Section 3 briefly describes our developed method
for determining a global spectrum of TIDs. The method is used to
analyze the data from the international GPS network for 10 days
with different levels of geomagnetic disturbance -- Section 4.
Results obtained are discussed in Section 5 , and compared with
available published data.

\section{General description and geometry of the experiment}
\label{sec:dis}

This study is based on using the data from the global GPS network
of receiving stations available via the Internet. Figure 1
presents the geometry of the global GPS network used in this
paper when analyzing the mean amplitude spectra of total electron
content (TEC) disturbances. For some reasons, slightly differing
sets of stations were chosen for different events which were
analyzed; however, the geometry of the experiment for all events
was virtually identical. We do not present here the coordinates of
the stations for reasons of space. This information may be
obtained at  the electronic address
http://lox.ucsd.edu/cgi-bin/allCoords.cgi?.

As is evident from Fig.~1, the set of stations selected from the
part of the global GPS network available to us, covers rather
densely North America and Europe, and much less densely Asia. The
number of stations on the Pacific and Atlantic oceans is smaller.
However, such coverage of the terrestrial surface makes it
possible even today to solve the problem of a global detection of
disturbances with an as yet unprecedented spatial accumulation.
This ensures a number of statistically independent series by two
orders of magnitude larger as a minimum than could be realized by
recording UHF signals from geostationary satellites \citep{Dav80,
Afr94} or from first-generation low-orbit navigation TRANSIT
satellites \citep{Eva83, Oga87}. Thus, in the western hemisphere
the corresponding number of stations can, already today, be as
large as 500, and the number of beams to satellites can be no
less than 2000-3000.

We carried out an analysis of the data for a set of from 100 to
300 GPS stations and for 10 days from the time interval
1998-2000, with a different level of geomagnetic disturbance
($Dst$ from -13 to -321 nT; $Kp$-index from 3 to 9). Table~1
presents information about day numbers, the number of the
stations used m, and extreme values of $Dst_{min}$ and
$Kp_{max}$. A total amount of the GPS data exceeds
$5\times10^{7}$ 30-s observations.

\section{Determining the mean (global) power spectrum of TEC variations
and its parameters from GPS data}
\label{sec:next5}

Below, we give a brief outline of our developed method for
estimating the mean (global) power spectrum of TEC variations
caused by ionospheric irregularities of different scales, on the
basis of processing the data from the international network of
two-frequency multichannel receivers of the GPS navigation
system. With the purpose of improving the statistical reliability
of the data, we used the global spatial averaging technique for
spectra within the framework of a novel GLOBDET technology
\citep{Afr2000a}. The method implies using an appropriate
processing of TEC variations that are determined from the GPS
data, simultaneously for the entire set of GPS satellites (as
many as 5--10 satellites) "visible" during a given time interval,
at all stations of the global GPS network used in the analysis.

The standart GPS technology provides a means for wave
disturbances detecion based on phase measurements of TEC at each
of spaced two-frequency GPS receivers. A methods of reconstructing
TEC variations from measurements of the ionosphere-induced
additional increment of the group and phase delay of the
satellite radio signal was detailed and validated in a series of
publications (Hofmann-Wellenhof  et  al.,  1992; Afraimovich et
al., 1998, 2000b). We reproduce here only the final formula for
phase measurements

\begin{equation}
\label{GMS-eq-1}
             I_o=\frac{1}{40{.}308}\frac{f^2_1f^2_2}{f^2_1-f^2_2}
                           [(L_1\lambda_1-L_2\lambda_2)+const+nL],
\end{equation}
where $L_1\lambda_1$ and $L_2\lambda_2$ are additional paths of
the radio signal caused by the phase delay in the
ionosphere,~(m); $L_1$ and $L_2$ represent the number of phase
rotations at the frequencies $f_1$ and $f_2$; $\lambda_1$ and
$\lambda_2$ stand for the corresponding wavelengths,~(m); $const$
is the unknown initial phase ambiguity,~(m); and $nL$~ are errors
in determining the phase path,~(m).

Phase measurements in the GPS can be made with a high degree of
accuracy corresponding to the error of TEC determination of at
least $10^{14}$~m${}^{-2}$ when averaged on a 30-second time
interval, with some uncertainty of the initial value of TEC,
however \citep{Hof92}. This makes possible detecting ionization
irregularities and wave processes in the ionosphere over a wide
range of amplitudes (up to $10^{-4}$ of the diurnal TEC
variation) and periods (from 24 hours to 5 min). The unit of TEC
$TECU$, which is equal to $10^{16}$~m${}^{-2}$~ and is commonly
accepted in the literature, will be used in the following.

Primary data include series of "oblique" values of TEC $I_o(t)$, as
well as the corresponding series of elevations $\theta(t)$ and
azimuths $\alpha(t)$ along LOS to the satellite calculated using
our developed CONVTEC program which converts the GPS system
standard RINEX-files on the INTERNET \citep{Gur93}.

Unfortunately, for most stations of the global GPS network, the
data are provided by the Internet at time intervals of 30 s,
which bounds the TEC variation period below by about 1 min.

A calculation of a single spectrum of TEC variations involves
using continuous series of $I_o(t)$ series of a duration of no less
than 2{.}5 hours, thus enabling us to obtain the number of counts
equal to 256 that is convenient for the algorithm of fast Fourier
transform (FFT) used in this study. To obtain a longer series of
512 counts requires a time interval no less than 5 hours long,
which is impracticable because of the limitations of the geometry
of experiment with the GPS satellites. This bounds the range of
periods analyzed by us above by about 120 min.

To exclude the variations of the regular ionosphere, as well as
trends introduced by the motion of the satellite, we employ the
procedure of removing the linear trend by preliminarily smoothing
the initial series with a selected time window of a duration of
about 60 min. This procedure reduces greatly the amplitude of
low-frequency components in the range of periods analyzed, but
this does not affect the qualitative results derived from
analyzing the spectrum below.

Series of the values of elevations $\theta(t)$ and azimuths
$\alpha(t)$ of the beam to the satellite were used to determine
the coordinates of subionospheric points, and to convert the
"oblique" TEC $I_{0}(t)$ to the corresponding value of the
"vertical" TEC by employing the technique reported by Klobuchar
(1986)

\begin{equation}
\label{GMS-eq-02} I = I_o \times cos
\left[arcsin\left(\frac{R_z}{R_z + h_{max}}cos\theta\right)
\right],
\end{equation}

where $R_{z}$ is the Earth's radius, and $h_{max}$=300 km is the
height of the $F_{2}$-layer maximum. All results in this study
were obtained for elevations $\theta(t)$ larger than
30$^\circ$.

By considering an example of the magnetically quiet and
magnetically disturbed ionosphere over the Millstone Hill
incoherent scatter facility - MHR (geographical coordinates
$42{.}61^\circ$N, $288{.}5^\circ$E), we describe briefly the
sequence of data processing procedures. Fig.~2a gives an example
of a typical weakly disturbed variation in "vertical " TEC $I(t)$
for station WES2 (satellite number PRN17) on July 15, 2000 for
the time interval 17{:}00-19{:}00 UT, preceding the onset of a
geomagnetic disturbance near the MHR over the territory with the
coordinates inside the rectangle $30-50^\circ$N, $270-290^\circ$E.
For this same series, Fig.~2b presents the $dI(t)$ variations
that were filtered out from the $I(t)$ series by removing the
trend with a 60-min window.

The logarithmic power spectrum $lg S^{2}(F)$ of the $dI(t)$ series
(Fig.~2b), obtained by using a standard FFT procedure, is
presented in panel c). Boldface letters and dots along the
abscissa axis on panel c (as well as d, g and h) indicate the
frequency ranges of medium-scale (MS) and small-scale (SS)
irregularities.

Incoherent summation of the partial power spectra $lg S^{2}(F)_i$
of different LOS was performed by the formula

\begin{equation}
\label{GMS-eq-03} \langle lg S^{2}(f)\rangle =
\sum^n_{i=1}lgS^{2}(f)_i,
\end{equation}

where $i$ is the number of LOS; $i=$ 1, 2, ... $n$.

The result derived from combining the spectra $\langle lg
S^{2}(f)\rangle$ for 16 LOS of 10 GPS stations located in the
above-mentioned MHR region is shown in Fig.~2d by a thick line.

For comparing the spectra for the quiet and disturbed days, the
thin line in Fig.~2d plots a global spectrum for the quiet day of
July 29, 1999 (a maximum deviation of $Dst$=-4 nT) obtained in a
similar manner for the time interval 11{:}00-13{:}30 UT by
averaging over n = 309 LOS of 161 stations of the global network,
Fig.~1, which are relatively uniformly distributed in the western
and eastern hemispheres within $30-70^\circ$N latitudes. Values
of the slope $k$ of the power spectrum are shown at the spectra.

As a consequence of the statistical independence of partial
spectra, the signal/noise ratio, when the average spectrum is
calculated, increases due to incoherent accumulation at least by
a factor of $\sqrt{n}$, where $n$ is the number of LOS. This is
confirmed by a comparison of the resulting sum of the $\langle lg
S^{2}(f)\rangle$, Fig.~2d, with the partial spectrum $lg
S^{2}(F)$, Fig.~2c.

It should be noted that the spectrum that is calculated directly
from $dI(t)$ variations is a distorted spectrum of irregularities
as a consequence of the Doppler shift effect of the TID angular
frequency \citep{Afr98}

\begin{equation}
\label{GMS-eq-4} \Omega=\Omega_0 - \vec{K}\vec{\omega},
\end{equation}

where $\vec{K}$ and $\vec{\omega}$ are, respectively, the TID
angular vector and the vector of displacement of the
subionospheric point at the selected height in the ionosphere
caused by the motion of the GPS satellite; $\Omega_0$ is the
initial value of the TID angular frequency.

The bulk of information about time spectra of different-scale
ionospheric irregularities, including TIDs, was obtained through
transionospheric soundings using signals from geostationary
satellites \citep{Dav80, Afr94}. In this case the velocity
$\omega$ of the beam to the satellite at the level of the
ionospheric $F_{2}$ region maximum is much smaller than the
velocity $V$ of TIDs, hence it can be neglected. For low-orbit
navigation satellites of the first-generation TRANSIT, on the
contrary, the velocity $\omega$ exceeds substantially the velocity
$V$ of TIDs; therefore, measurements are interpreted in terms of
one-dimensional spatial spectra \citep{Eva83}. In the case of the
GPS, the velocities can be identical, which will cause the
spectral line to be shifted toward the positive or negative sides.

The resulting value of the frequency can change sign if the
modulus of frequency shift $|\vec{K} \vec{\omega}|$ exceeds the
value of $\Omega$. This means that in this case the point at
which LOS traverses the main maximum of ionization moves faster
than the TID wave, and in the interferometer's frame of reference
the travelling direction of the equiphase line is reversed with
respect to that in the ionosphere. Such a situation, however, can
be of very infrequent occurrence because the velocity $\omega$
(usually not higher than 50-70 m/s when $h_{\rm{max}}$=300 km) is
distinctly lower than the mean value of the TID phase velocity.

However, as partial spectra are accumulated, which correspond to
all GPS satellites that are visible over a given time interval,
this effect will lead merely to a relatively uniform smearing of
spectral lines because the sign and magnitude of the frequency
shift are different for separate LOS. Thus an averaging over a
large number of LOS makes it possible to obtain estimates of
average spectra.

As is evident from Fig.~2d, the spectrum of a quiet day
corresponds reasonably well to a theoretical power spectrum of
ionospheric irregularities with a slope of about $k$=-2{.}5, and it
can therefore be used as a reference power spectrum. This result
is consistent with published estimates of TID spectrum
characteristics obtained in vertical- \citep{Kali88},
oblique-incidence \citep{Gaj83} and transionospheric radio
soundings \citep{Afr94}. The TEC fluctuation scale $M$ in the MS
range and $C$ in the SS range does not exceed in this case the
values 0{.}4 and 0{.}007 TECU, respectively.

When comparing the average spectra of TEC variations from July
15, 2000 for the time interval 17{:}00-19{:}00 UT with the
spectrum from the quiet day of July 29, 1999, one can notice an
order-of-magnitude excess of the TEC disturbance level throughout
the spectrum, with the value of the slope $k$=-2{.}56 remaining
the same. However, there is also a clear disproportionate (by
1{.}5 order of magnitude) increase in TEC variation intensity in
the MS range.

Still more drastic changes of the ionospheric irregularity
spectrum occurred over the same region just one hour later.
Figure 2e presents the time dependence of the disturbed value of
the "vertical" TEC $I(t)$ for station ALGO (satellite number
PRN21) for July 15, 2000, for the time interval 20{:}00-22{:}30
UT. For the same series, Fig.~2f plots the $dI(t)$ variations
that were filtered out from the $I(t)$ series by removing the
trend with a 60-min window. As is apparent from Fig.~2a, and from
the corresponding $lg S^{2}(F)$ spectrum, Fig.~2g, the TEC
variations increased in power as a minimum by 2 orders of
magnitude as against the time interval 17{:}00-19{:}00 UT
(Fig.~2b and 2c). Besides, there was an abrupt change in the
spectrum slope $k$ =-0{.}85, which is indicative of a
disproportionate increase in irregularity intensity in the MS and
SS parts of the spectrum. The TEC fluctuation scale $M$ in the MS
range and $C$ in the SS range exceeds in this case the values
4{.}27 and 0{.}5 TECU, respectively.

The result derived from combining the $\langle lg S^{2}(f)\rangle$
spectra for 7 LOS is shown in Fig.~2h by a thick line. The
spectrum has a power-law character, yet with the mean slope
$k$=-1{.}85, which differs markedly from the value of $k$ for the
magnetically quiet day. The mean intensity $M$ of the
irregularities of the medium-scale part MS increased two orders
of magnitude, and the intensity $C$ of the small-scale part
increased immediately by 3 orders of magnitude as compared with
the level of the magnetically quiet day.

As the chief goal of this paper is to obtain the mean
characteristics of the TID intensity, in the discussion to follow
we shall be using only the above-mentioned parameters of the
spectrum, $k$ and $M$.

\subsection{Geomagnetic control of the TID spectrum}

The dependencies of the TID variation intensities $M(t)$ presented
below were obtained by calculating global spectra for all days
listed in Table~1, with the number of stations $m$ for time
intervals of a duration of 2{.}5 hours with a 1-hour shift, and by
a subsequent integration of the spectral density in the range of
20--60 min periods (see Fig.~2, where boldface dots along the
abscissa axis indicate the MS interval). The integration result
is the value of $M$ which is equal to the standard deviation of
TEC variations in the specified range of periods and is measured
in TECU units. Corresponding data are presented in Figs.~3, 4 and
5, and statistical estimates are listed in Table~1.

\subsubsection{Characteristics of the TID spectrum as a function of the
universal time UT}

The data from the magnetically quiet day of July 29, 2000,
characterized by a low level of geomagnetic activity and by a
reference power spectrum (see Fig.~2d, 2h - thin line), are used
here in comparison with characteristics of the TID spectrum
during geomagnetic disturbances (line 3 in Table~1). Figure 3e
(thick line) and Fig.~3f (dashes) plot, as a function of the
universal time UT, the $Dst$-variations of the geomagnetic field
and the standard deviation of the TEC variations $M(t)$ in the
range of 20--60 min periods for this day.

As is evident from the figure, shallow, slow $Dst$-variations over
the course of that day were accompanied by slow, small-amplitude
fluctuations caused by TIDs; the mean value $M$ for that day did
not exceed 0{.}16 TECU. Similar results were also obtained for the
other magnetically quiet day of January 9, 2000 (Fig.~3b - thick
line, and Fig.~3c - dashes; line 4 in Table~1).

Let us now consider, for the sake of contrast, a global
ionospheric response to a major magnetic storm of April 6--7,
2000, characterized by a maximum amplitude of $Dst$-variations as
large as -321 nT (Fig.~3e - thin line; line 5 in Table~1). A
maximum value of the $Kp$ index (Fig.~3d) for this storm was as
high as 8. Until about 19{:}00 UT on April 6, the $Dst$-variations
varied within a very narrow range, and were close to 0 nT. After
that, the value of $Dst$ began to decrease rapidly; after 19{:}00
UT it reaches the value -129 nT, and continued decreasing right
down to -321 nT.

Fig.~3f plots the dependence $M(t)$ of the standard deviation of
TEC variations in the range of 20--60 min periods (thick line),
and the inverted dependence of the time derivative $d(Dst)/dt$
(relative units; thin line). The derivative $d(Dst)/dt$ was
obtained from the dependence $Dst(t)$ (Fig.~3e) that was smoothed
with a 7-hour time window. As is evident from the figure, an
increase of the level of magnetic disturbance is accompanied by a
gradual increase in total intensity of TIDs; however, it
correlates not with the absolute level of $Dst$ but with the
value of the time derivative $d(Dst)/dt$ (the correlation
coefficient $r$ in this case is -0{.}84). A maximum amplitude
$M_{max}$ = 1{.}07 TECU, shown by the arrow in Fig.~3f, exceeds
one order of magnitude as a minimum the corresponding value for
the magnetically quiet day of July 29, 1999 (Fig.~3f - dashes).

Similar results were also obtained for other magnetic storms from
July 15--16, 2000 (Fig.~3a-c), August 26--27, and September
24--25, 1998 (Fig.~4), yet a maximum value $M_{max}$ in these
cases did not exceed 0{.}67, 0{.}32 and 0{.}42 TECU, respectively
(lines 6, 1 and 2 in Table~1).

The delay $\tau$ (of about 2 hours) of the increase in TEC
intensity with respect to rapid changes in magnetic field
strength is easy to explain by taking into consideration that the
greatest contribution in a global averaging of TID spectra is
made by the mid-latitude chain of GPS stations. This chain is at
about 2000 km from the southern boundary of the auroral source of
TIDs which is produced during geomagnetic disturbances. TIDs that
are generated once this source is produced travel equatorward with
the velocity of order 300-400 m/s \citep{Fra73, Mae80, Hun82,
Hay87, Ma98, Hoc96, Ho98, Bal99, Hal99, Afr2000b}.

\section{Characteristics of the TID spectrum as a function of the
local time LT}
\label{sec:end}

For studying the diurnal dependence of TID spectrum
characteristics, we carried out an averaging of the spectra with
due regard for the local time LT for each GPS station. In doing
so, it should be taken into consideration that as a consequence
of the nonuniform distribution of the stations, the contribution
of the mid-latitude stations in North America and, to a lesser
extent, in Europe is predominant (see Fig.~1).

Figure 5 plots the diurnal LT-dependencies of the slope index $k$
of the power spectrum of TIDs -- a), and of the standard
deviation of the TEC variations $M(t)$ in the range of 20--60 min
periods -- b) for the magnetically quiet day of July 29, 1999. As
is evident from the figure, the value of the slope index
$k$=-2{.}5 remains virtually unchanged over the course of that
day, as does the mean value $M(t)$ which does not exceed 0{.}15
TECU.

However, for the other magnetically quiet day, January 9, 2000
(Fig.~5c, d), one can notice a conspicuous diurnal dependence of
both the slope index $k(t)$ and the TID intensity $M(t)$. Also, a
maximum value of $M(t)$ is as high as 0{.}8 TECU around noon, and
a maximum slope index $k$ is as high as -2{.}7.

Thus, according to our data, the power spectra of the daytime TEC
variations in the range of 20--60 min periods under quiet
conditions have a power-law form with the slope index $k$=-2{.}5.
With the increasing level of geomagnetic disturbance, there is an
increase in total intensity of TIDs, with a concurrent kink of
the spectrum caused by an increase in fluctuation intensity in
the range of 20--60 min. The TEC variation amplitude is smaller at
night than during the daytime, and the spectrum decreases in
slope, which is indicative of a disproportionate growth of the
amplitude of the small-scale part of the spectrum.

The above-mentioned characteristic features of the diurnal
variation of these parameters are most pronounced during a major
magnetic storm of April 6, 2000 - Fig.~5e, f and, to a lesser
extent, during the magnetic storm of July 15, 2000 - Fig.~5g, h).

The peculiarities of the diurnal variation in TID intensity
pointed out above are consistent with evidence acquired using
signals from geostationary satellites \citep{Jac95, Oli97, Afr99}.

The characteristics of the spectra which we have obtained are in
reasonably good agreement with a number of reported results,
despite the fact that the published data show a large scatter in
estimates of the slope $k$ (as well as of the amplitude scale $M$
of temporal and spatial spectra).

One-dimensional spatial spectra can be obtained through direct
measurements of variations in local electron density along the
satellite path; however, published data mostly refer to the
equatorial or polar regions. Thus, by investigating the
equatorial F region of the ionosphere simultaneously through
in-situ measurements by satellites AE-E and radio probing using
signals from the geostationary Wideband satellite at 137 and 378
MHz frequencies, Livingston et al. (1981) found that the
one-dimensional spatial spectrum in the range of scales of 10-100
km has a power character with the slope index $k$ of about -2.

Rocket measurements of electron densities in the $F$ region of the
auroral ionosphere made concurrently with incoherent scatter
radar measurements and radio probing using signals from the
geostationary Wideband satellite at Chatanika \citep{Kel80} showed
that the corresponding slope index $k$ of the one-dimensional
spatial spectrum for the range of 0{.}1-200 km scales lies within
-1{.}2-1{.}8.

For 500 km altitude, estimates of the slope of the
one-dimensional spatial spectrum were obtained from in-situ
measurements aboard the "Cosmos-900" satellite by Gdalevich et
al. (1980). While for the equatorial and high-latitude ionosphere
the values of $k$ were of order -1{.}2, the spectrum for the
mid-latitudes showed a kink; in the range of 30-150 km scales, $k$
= -3-4, and it decreases to -1{.}0 for irregularities smaller than
30 km.

An alternative possibility involves transferring the temporal
spectrum to the spatial region provided that the traveling
inhomogeneous structure is "frozen-in". In this case the form of
the spatial spectrum is identical to that of the temporal
spectrum. Such an approach in processing measurements of the
frequency Doppler shift at oblique-incidence soundings was used,
in particular, by Gajlit et al. (1983); a corresponding estimate
of the index k for ionospheric irregularities with the size from
a few tens to several hundred kilometers was obtained by these
authors in the range of -3{.}8-4{.}6. Furthermore, the index $k$
for the spectra of the frequency Doppler shift varied within
-0{.}8-1{.}6. Similar results in identical Doppler measurements
were obtained by Kaliev et al. (1988); the $k$ from the data on
average spectra was -2 both during the daytime and at night.

Afraimovich et al. (1994) investigated the spectral properties of
medium-scale TIDs on the basis of analyzing power spectra of TEC
variations obtained by measuring the polarization of the signal
from geostationary satellite ETS-2 at 136 MHz frequency near
Irkutsk ($52^\circ$ N, $102^\circ$ E). For three seasons of 1990,
temporal spectra of TEC variations, averaged over 10 days, were
obtained. In the low-frequency range (periods from 100 to 20 min)
the daytime variation spectra have a power-law form, with the
slope index - 2.5, while in the high-frequency range (periods of
20--10 min) they are with the index - 6; at night, the index is -
4 throughout the range of periods under consideration.

\section{Discussion and conclusion}
\label{GMS-sect-5}

The main results of this study may be summarized as follows:

\begin{enumerate}

\item Our findings bear witness to the determining role of geomagnetic
disturbances in the formation of the spectrum of traveling
ionospheric disturbances. This conclusion is based on
substantially more extensive (than obtained earlier) statistical
material, spans periods with a different level of geomagnetic
disturbance, and has a global character. The analysis has been
made for a set of 100 to 300 GPS stations, and for 10 days with a
different level of geomagnetic activity ($Dst$ from 0 to -350 nT;
the $Kp$ index from 3 to 9).

\item It was found that power spectra of daytime TEC variations in the
range of 20--60 min periods under quiet conditions have a
power-law form, with the slope index $k$=-2{.}5. With an increase
of the level of magnetic disturbance, there is an increase in
total intensity of TIDs, with a concurrent kink of the spectrum
caused by an increase in oscillation intensity in the range of
20--60 min. The TEC variation amplitude is found to be smaller at
night than during the daytime, and the spectrum decreases in
slope, which is indicative of a disproportionate increase in the
amplitude of the small--scale part of the spectrum.

\item It was found that an increase in the level of geomagnetic
activity is accompanied by an increase in total intensity of TEC;
however, it correlates not with the absolute level of $Dst$, but
with the value of the time derivative of $Dst$ (a maximum
correlation coefficient reaches -0{.}94).

\item The delay of the TID response of the order of 2 hours
is consistent with the view that TIDs are generated in auroral
regions, and propagate equatorward with the velocity of about
300--400 m/s.
\end{enumerate}

\begin{acknowledgements}
The author is grateful to N.~N.~Klimov and E.~A.~Ponomarev for
their encouraging interest in this study and active participation
in discussions. Thanks are also due V.~G.~Mikhalkovsky for his
assistance in preparing the English version of the \TeX
manuscript. This work was done with support from the Russian
Foundation for Basic Research (grant 99-05-64753) and from RFBR
grant of leading scientific schools of the Russian Federation
00-15-98509.
\end{acknowledgements}

\end{document}